\documentclass[twocolumn,showpacs,preprintnumbers,amsmath,amssymb,floatfix]{revtex4}
\usepackage{amsmath,amsfonts,amssymb,amsthm,epsfig,epstopdf,url,array}
\usepackage{graphicx}
\usepackage{dcolumn}
\usepackage{bm}

\newcommand{\beq}{\begin{equation}}
\newcommand{\eeq}{\end{equation}}
\newcommand{\bey}{\begin{eqnarray}}
\newcommand{\eey}{\end{eqnarray}}

\begin{document}
\title{ Neutron Stars : A Comparative Study }

\author{ Mehedi Kalam}
\email{kalam@iucaa.ernet.in} \affiliation{Department of
Physics, Aliah University, IIA/27, New Town,  Kolkata -
700156, India}

\author{Sk. Monowar Hossein}
\email{hossein@iucaa.ernet.in} \affiliation{Department of
Mathematics, Aliah University, IIA/27, New Town,  Kolkata -
700156, India}

\author{Sajahan Molla}
\email{sajahan.phy@gmail.com} \affiliation{Department of Physics,
Aliah University, IIA/27, New Town,  Kolkata -156, India}

\date{\today}

\begin{abstract}
The inner structure of neutron star is considered from theoretical point of view and is compared with the observed data.  We have proposed a form of an equation of state 
relating pressure with matter density which indicates the stiff equation of state of neutron stars. From our study we have calculated mass(M), compactness(u) and surface red-shift($Z_{s}$) 
for the neutron stars namely  PSR J1614-2230, PSR J1903+327, Cen X-3, SMC X-1, Vela X-1, Her X-1  and compared with the recent observational 
data. We have also indicated the possible radii of the different stars which needs further study. Finally we have examined the stability for such type of theoretical structure.
\end{abstract}


\maketitle
\section{Introduction}
Study of compact objects like star takes much attention to the astrophysicists for the last few Decades. It plays the crucial role like a bridge among astrophysics, 
nuclear physics and particle physics. One of the possible existing compact objects in our universe is neutron star (or strange star). Neutron stars are the most 
acceptable and known compact objects for the study of dense matter physics. In general, neutron stars are composed 
almost entirely of neutrons, while strange stars are made from strange quark matter(SQM). If we consider strange quark matter is stable then strange stars may be formed 
during supernova explosions. As a consequence of this neutron stars can be converted to strange stars by a number of different mechanisms such as: 
a) pressure-induced transformation to uds-quark matter
via ud-quark matter, b) sparking by high-energy neutrinos, c) triggering due to the intrusion of a quark nugget. All of these possibilities were described 
by C. Alcock, E. Farhi, and A. Olinto\citep{C. Alcock1986}. Scientific community beleive that a much more rapid cooling of SQM in strange stars takes place than neutron stars 
due to neutrino emitting weak interactions involving the quarks\citep{C. Alcock1986}. Therefore, a strange star was known as much colder than a neutron star of similar age. 
The most massive compact stars are entirely made of quarks, i.e. they are quark stars\citep{A. Drago2014,P. Haensel1986}. 
Famous scientist Edward Witten\citep{E. Witten1984}  concluded that ``If quark matter is stable, 
it is probably necessary to assume that ordinary neutron stars are really quark stars".
Recently observed neutron star or strange star have been detected in the Milky Way and only very few of them have been discovered in globular clusters. Some of the possible 
existing compact stars(neutron star) are PSR J1614-2230,  PSR J1903+327, Cen X-3, SMC X-1, Vela X-1, 
Her X-1\citep{P B Demorest2010,P. C. C. Freire2011,M L Rawls2011,M. J. Coe2013}. Many researcher 
studied\citep{Rahaman2012a,Kalam2012a,Hossein2012,Rahaman2012b,Kalam2012b,Kalam2013,Kalam2014a,Kalam2014b,Kalam2014c,Hossein2014,Lobo2006,Bronnikov2006,Egeland2007,Dymnikova2002}
 compact stars in various directions. Scientists used different techniques such as computational, observational or theoretical analysis to study  astrophysical objects. 
Since the uncertainty in the behavior of matter inside the compact stars ( ''normal matter for neutron stars" or
``strange quark matter for strange stars"), their physical
structure and some physical property can be obtained by applicable analytic solutions of Einstein’s
gravitational field equations. The Relativistic stellar model have been studied by Karl Schwarzschild in 1916\citep{K. Schwarzschild1916}, occasionally the first solution of 
Einstein’s field equation for the
interior of a compact object which are in hydrostatic equilibrium. In this regard, an important 
investigation was done by Tolman \citep{R. C. Tolman1939} in 1939. In his paper, he proposed eight evident analytical solutions of the field equations.
Clifford E. Rhoades, Jr. and Remo Ruffini\citep{Clifford E. Rhoades1974} showed that maximum possible mass of a neutron star can not be larger 
 than 3.2$M_{\odot}$. 
Buchdahl\citep{H. A. Buchdahl1959} also contributed an important study of
the fluid spheres by obtaining the famous bound to the
mass(M) - radius(R) ratio for stable general relativistic spheres, which is $\frac{2GM}{c^2R} \leq \frac{8}{9}$.\\
Inspite of considerable progress in current years, still there are some important features to be addressed about neutron stars. Observations of masses and radii 
confirms the theoretical model for white dwarf stars. However, due to the uncertainities in the equation of state of neutron stars, observations of masses and radii are 
used to test theories of neutron stars.\\
 Motivated by the above facts we are introducing a theoretical model of neutron stars which indicates the possible equation of state, determine the masses and radii of various
 neutron stars. The solution presented in this article satisfies all the energy conditions, TOV-equation, and Buchdahl mass-radius relation\citep{H. A. Buchdahl1959} 
 which is required for the stability of the stars. 
 It also satisfies other stability condition like speed of sound ($< 1$) inside the star and adiabetic index ($ > \frac{4}{3} $) in radial adiabatic perturbations. 
 In this article we are able to give an equation of state(EOS) relating pressure and matter density of 
 the star which is also a support for stiff equation of state for neutron star. 
 Interestingly, the star masses calculated from our model is well matched  with the observed stars masses, which confirm the validity of our model for real case. \\ 
 In Sec II, we have discuss the interior spacetime and behavior of the star. In Sec. III, we have studied some special features of the star namely, TOV equation, 
 Energy conditions, Stability, Mass-radius relation, Compactness and Surface redshift in different sub-sections. In Sec. IV, The article concluded
with a short discussion with numerical data and concluding remarks. Before we start it is worthy to mention that, we use geometric units $G = c = 1$ throughout the article.

\section{Interior Spacetime of the star}

In our stellar model, we consider a static and spherically symmetric matter distribution whose interior space time as:
\begin{eqnarray}
ds^2 = -e^{\nu(r)}~dt^2 + e^{\lambda(r)}~dr^2 + r^2 (d\theta^2 +sin^2 \theta d\phi^2)
\end{eqnarray}
 According to H. Heintzmann\citep{H. Heintzmann1969},
 \begin{eqnarray}
e^{\nu} &=&
A^2\left(1+ar^2\right)^{3}
\end{eqnarray}
and 
\begin{eqnarray}
e^{-\lambda} &=&
\left[1-\frac{3ar^2}{2} \left(\frac{1+C\left(1+4ar^2\right)^{-\frac{1}{2}}}{1+ar^2}\right) \right]
\end{eqnarray}
where A (dimensionless), C(dimensionless) and $a$(length$^{-2}$) are constants.\\
The matter within the star is to be considered as perfect fluid (locally) and consequently the energy momentum tensor for the fluid distribution can be 
written as 
\begin{equation}
 T^i_j = (\rho + p)v^iv_j - p \delta^i_j, \nonumber
\end{equation}
where $v^i$ is the four velocity,$\rho$ is the matter density and $p$ is the fluid pressure.



\begin{figure}[htbp]
\centering
\includegraphics[scale=.3]{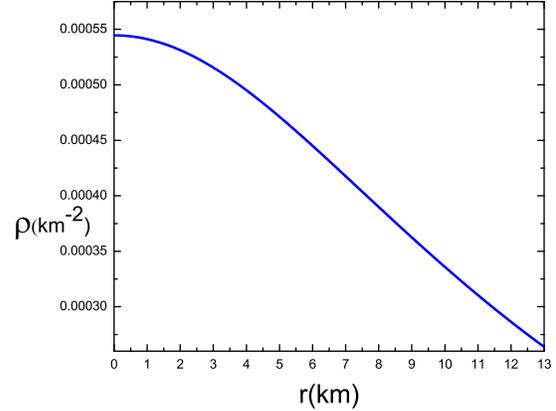}
\caption{Density($\rho$) - radius(r) relation at the steller interior (taking a=0.002, C=0.52) where density in unit of $km^{-2}$ and radius in unit of $km$. }
\label{fig:1}
\end{figure}

\begin{figure}[htbp]
\centering
\includegraphics[scale=.3]{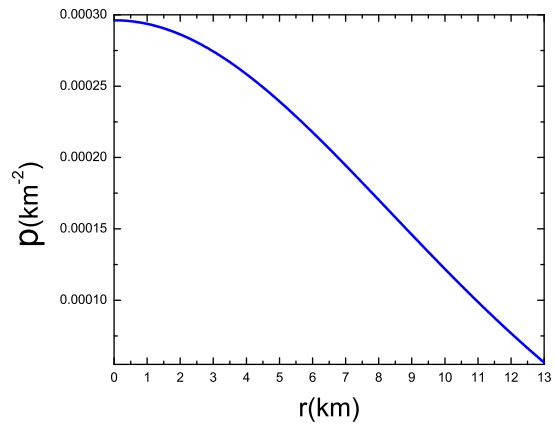}
\caption{Pressure($p$) - radius(r) relation at the steller interior (taking a=0.002, C=0.52) where pressure in unit of $km^{-2}$ and radius in unit of $km$.}
\label{fig:2}
\end{figure}
Solving the Einstein's field equation we get,
\begin{eqnarray}
\rho &=& \frac{3a\left(\sqrt{1+4ar^2}\left(3+13ar^2+4a^2r^4\right)+C\left(3+9ar^2\right)\right)}{16\pi\left(1+ar^2\right)^2\left(1+4ar^2\right)^{\frac{3}{2}}} ,\nonumber\\
\rho_0 &=& \rho(r=0) = \frac{3a\left(3+3C\right)}{16\pi}  ,\nonumber
\end{eqnarray}
and
\begin{eqnarray}
 p &=& \frac{-3a\left(3\sqrt{1+4ar^2}\left(-1+ar^2\right)+C+7aCr^2\right)}{16\pi\left(1+ar^2\right)^2\left(1+4ar^2\right)^{\frac{1}{2}}} ,\nonumber\\
p_0 &=& p(r=0) = \frac{3a\left(3-C\right)}{16\pi}  ,\nonumber
\end{eqnarray}
where $ \rho_0$ and $ p_0$ are the central density and central pressure of the star respectively.

We observe that, pressure and density are maximum at the centre and decreases
monotonically towards the boundary(FIG. 1 and FIG. 2). Therefore, they are well behaved in the interior of the stellar structure.

\section{Some special features}
\subsection{TOV equation}
The modified Tolman-Oppenheimer-Volkoff (TOV) equation describes the equilibrium condition for the compact star subject to
the gravitational($F_g$) and hydrostatic($F_h$) forces,
\begin{equation}
F_h+ F_g  = 0,\label{eq21}
\end{equation}
where,
\begin{eqnarray}
F_g &=& \frac{1}{2} \nu^\prime\left(\rho+p\right)\label{eq22}\\
F_h &=& \frac{dp}{dr} \label{eq23}
\end{eqnarray}
Therefore, the static equilibrium configurations do exist in the presence of gravitational and hydrostatic  forces (FIG 3).
\begin{figure}[htbp]
\centering
\includegraphics[scale=.3]{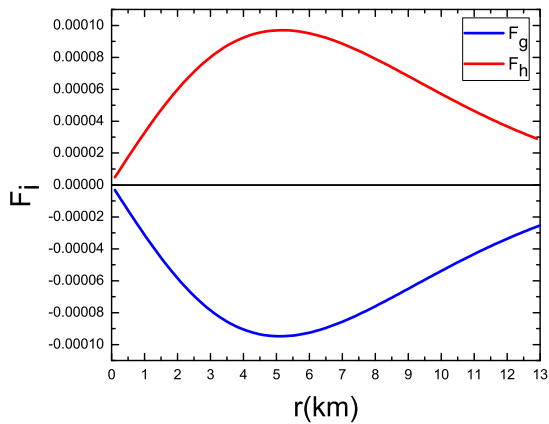}
\caption{Gravitational and hydrostatic forces at the stellar interior.}
\label{fig:3}
\end{figure}

\subsection{Energy conditions}
We verify whether all the energy conditions, like, null energy condition(NEC), weak energy condition(WEC), strong energy condition(SEC)
and dominant energy condition(DEC) are satisfied at the centre ($r=0$) of the star or not. For this purpose the following equations to be satisfied:\\

\begin{table*}[t]
\label{table:1}
\caption{Parameters for energy conditions}
\begin{tabular}{|l|l|l|l|l|l|l|}
\hline
 $\rho_{0}$ (km$^{-2}$) & $p_{0}$ (km$^{-2}$) & $\rho_{0}$+$p_{0}$ (km$^{-2}$)  & 3$p_{0}$+$\rho_{0}$ (km$^{-2}$)  \\
\hline
 0.000543987 & 0.000296278 & 0.000840265 & 0.001432821 \\
 \hline
\end{tabular}
\end{table*}

(i) NEC: $p_{0}+\rho_{0}\geq0$ ,\\
(ii) WEC: $p_{0}+\rho_{0}\geq0$  , $~~\rho_{0}\geq0$  ,\\
(iii) SEC: $p_{0}+\rho_{0}\geq0$  ,$~~~~3p_{0}+\rho_{0}\geq0$ ,\\
(iv) DEC: $\rho_{0} > |p_{0}| $.

From FIG. 1 and FIG. 2 we observe that all the energy conditions are satisfied (see Table:I).

\subsection{Stability}
For a physically acceptable model, one expects that the velocity of
sound should be less than light velocity$(c=1)$ i.e. within the range  $0 \leq  v^2=(\frac{dp}{d\rho})
\leq 1$ \citep{Herrera1992,Abreu2007}. \\

We plot the sound speed in FIG.4 and observe that it satisfy the inequality $0\leq v^2 \leq 1$ everywhere within the stellar object. This indicates the stability of the model.
\begin{figure}[htbp]

  \centering
        \includegraphics[scale=.3]{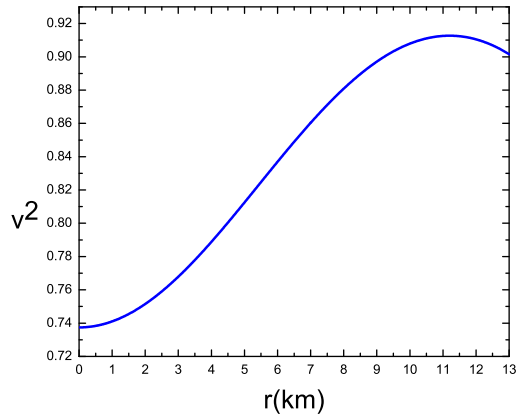}
       \caption{ Sound speed - radius relationship at the stellar interior(assuming a=0.002, C=0.52).}
   \label{fig:4}
\end{figure}

We also investigate the dynamical stability of the stellar model against the infinitesimal radial adiabatic perturbation which was introduced by 
S. Chandrasekhar\citep{S. Chandrasekhar1964}. Later this stability condition was devoloped and applied to astrophysical cases by 
J. M. Bardeen, K. S. Thorne, and D. W. Meltzer\citep{J. M. Bardeen1966}, H. Knutsen\citep{H. Knutsen1988}, M. K. Mak and T. Harko\citep{T. Harko2013}, gradually. 
The adiabatic index($\gamma$) is defined as\\
\begin{equation}
 \gamma = \frac{\rho c^2+p}{p} \frac{1}{c^2} \frac{dp}{d\rho}.
\end{equation}
The above equation can be written as (assume c=1)
\begin{equation}
 \gamma = \frac{\rho+p}{p} \frac{dp}{d\rho}
\end{equation}
Since $\gamma$ should be $> \frac{4}{3}$ everywhere within the isotropic stable star, we plot the adiabetic index for our model (FIG. 5) and 
observe that $\gamma > \frac{4}{3}$ everywhere within the star.  Therefore this stellar model is stable against the radial adiabatic infinitesimal perturbations.\\

\begin{figure}[htbp]

  \centering
        \includegraphics[scale=.3]{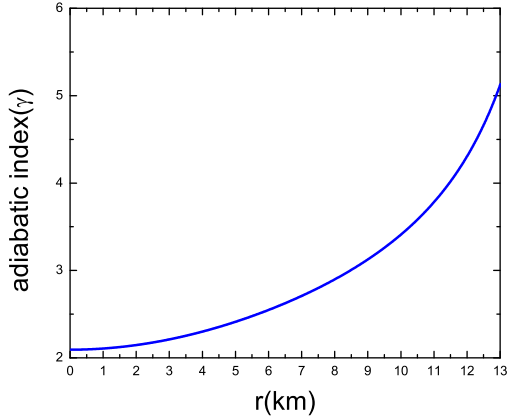}
       \caption{ Relation between adiabatic index $\gamma $ with the radius of the star at the stellar interior.}
   \label{fig:5}
\end{figure}

\subsection{Mass-Radius relation and Surface redshift}
 According to Buchdahl \citep{H. A. Buchdahl1959}, for a
static spherically symmetric perfect fluid sphere, maximum allowable mass-radius
ratio should be $\frac{ Mass}{Radius} < \frac{4}{9}$. Mak.\citep{Mak2001} also proposed more simplified expression.
In our model, the gravitational mass (M) in terms of the energy density ($\rho$) can be expressed as
\begin{equation}
\label{eq34}
 M=4\pi\int^{b}_{0} \rho~~ r^2 dr =
 \frac{3ab^3\left(C+\sqrt{1+4ab^2}\right)}{4\left(1+ab^2\right)\sqrt{1+4ab^2}}
\end{equation}
where $b$ is the radius of the star.\\
 The compactness(u) of the star should be
\begin{equation}
\label{eq35} u= \frac{ M(b)} {b}=
 \frac{3ab^2\left(C+\sqrt{1+4ab^2}\right)}{4\left(1+ab^2\right)\sqrt{1+4ab^2}}.
\end{equation}
The nature of the Mass and Compactness of the star are shown in FIG. 6 and FIG. 7.

\begin{figure}[htbp]
\centering
\includegraphics[scale=.3]{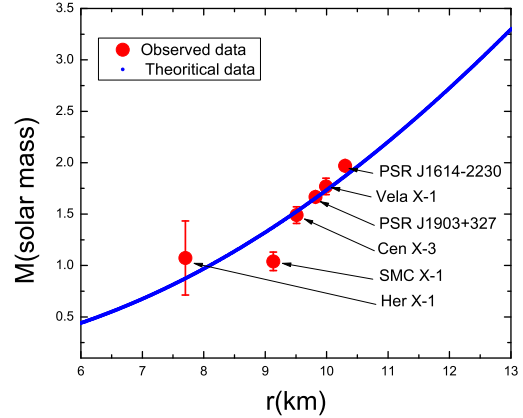}
\caption{The predicted mass(M) - radius(r) relation with observed data at the stellar interior (assuming a = 0.002, C = 0.52).}
\label{fig:6}
\end{figure}

\begin{figure}[htbp]
\centering
\includegraphics[scale=.3]{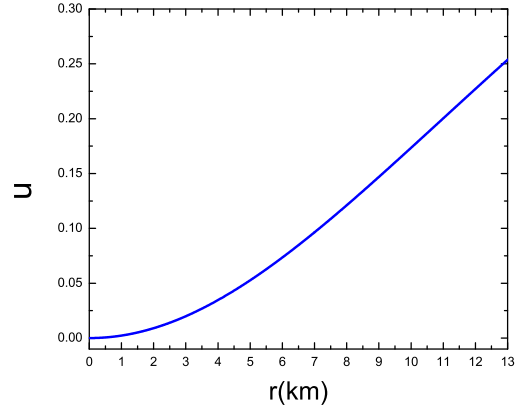}
\caption{Variation of the Compactness(u) at the stellar interior (assuming a = 0.002, C = 0.52).}
\label{fig:7}
\end{figure}

The surface redshift ($Z_s$) corresponding to the above
compactness ($u$) can be written as
\begin{equation}
\label{eq36} 1+Z_s= \left[ 1-(2 u )\right]^{-\frac{1}{2}} ,
\end{equation}
where
\begin{equation}
\label{eq37} Z_s= \frac{1}{\sqrt{1-\frac{3ab^2\left(C+\sqrt{1+4ab^2}\right)}{2\left(1+ab^2\right)\sqrt{1+4ab^2}}}}-1
\end{equation}
Therefore, from FIG. 8, the maximum surface redshift for the isotropic neutron stars  of
different radii can be obtain.  Mass, compactness, surface redshift of the neutron stars are evaluated from the above equations ( 9, 10, 12) and a comparative analysis
 is done in Table II.

\begin{figure}[htbp]
    \centering
        \includegraphics[scale=.3]{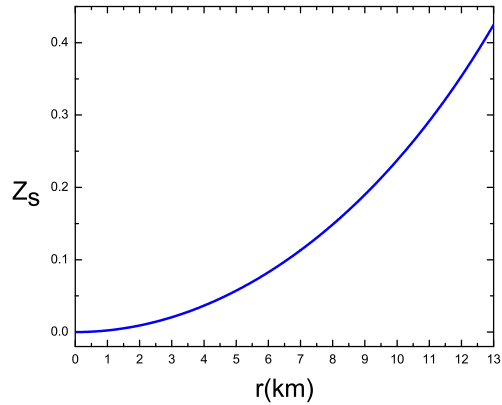}
        \caption{ Variation of the red-shift function at the stellar interior (assuming a = 0.002, C = 0.52).}
    \label{fig:8}
\end{figure}

\begin{table*}[t]
\label{table:1}
\caption{Evaluated parameters for Neutron Stars}
 \begin{tabular}{|c| c| c| c| c| c|}
 \hline
 Star & Radius(in km) & Observed Mass($M_{\odot}$) & Mass from Model($M_{\odot}$) & Redshift & Compactness \\ [0.3ex]
 \hline
 PSR J1614-2230 & 10.3 & 1.97 $\pm$ 0.04 & 1.86932 & 0.252915 & 0.2677\\
 \hline
 Vela X-1 & 9.99 & 1.77 $\pm$ 0.08 & 1.73007 & 0.23689 & 0.2554\\
 \hline
 PSR J1903+327 & 9.82 & 1.667 $\pm$ 0.021 & 1.65602 & 0.228382 & 0.2487\\
 \hline
 Cen X-3 & 9.51 & 1.49 $\pm$ 0.08 & 1.52525 & 0.213363 & 0.2366\\
 \hline
 SMC X-1 & 9.13 & 1.04 $\pm$ 0.09 & 1.37254 & 0.195798 & 0.2217\\
 \hline
  Her X-1 & 7.7 & 1.073 $\pm$ 0.36 & 0.874387 & 0.137476 & 0.1675\\ [0.3ex]
 \hline
\end{tabular}
\end{table*}

\begin{figure}[htbp]
\centering
\includegraphics[scale=.3]{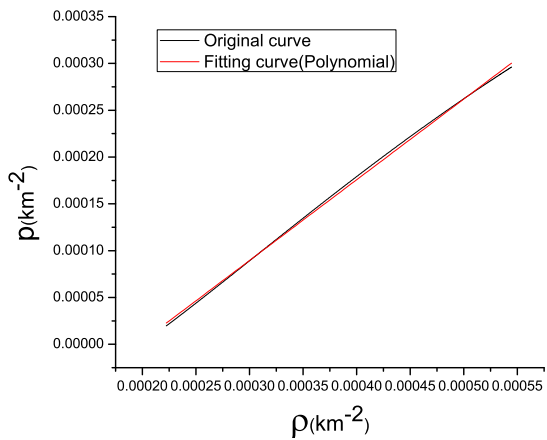}
\caption{Polynomial approximation of Pressure ($p$) - density ($\rho$) relation.}
\label{fig:9}
\end{figure}

\begin{figure}[htbp]
\centering
\includegraphics[scale=.3]{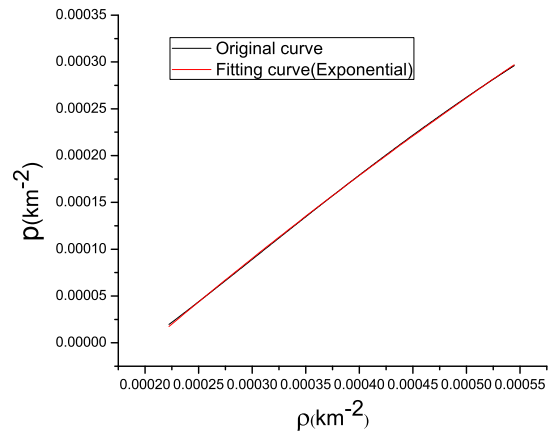}
\caption{Exponential approximation of Pressure ($p$) - density ($\rho$) relation.}
\label{fig:10}
\end{figure}


\section{Discussion and Concluding remarks}
For a binary system, Jacoby et al.\cite{Jacoby2005} and Verbiest et al. \cite{Verbiest2008} used the detection of Shapiro delay to measure the masses of 
both the neutron star and its binary companion. Using the same approach, Demorest et al.\cite{Demorest2010} has performed radio timing observations for the binary
millisecond pulsar PSR J1614-2230. The measured mass for the above pulsar in \cite{Demorest2010} is $1.97 \pm 0.04 M_{\odot} $ up to the highest precession reported yet.\\
With the help of Arecibo and Green Bank radio timing observations and considering the relativistic Shapiro delay, Freire et al\cite{P. C. C. Freire2011} obtained new constraints
on the mass of the pulsar and its companion and determine the accurate mass for PSR J1903+327 as $1.667 \pm 0.021 M_{\odot} $ through a detailed analysis. Rowls et al.\cite{M L Rawls2011}
 have determine the mass of neutron stars such as (Vela X-1, Cen X-3, SMC X-1, Her X-1) in eclipsing X-ray pulsar binaries. Their measured values are $1.77 \pm 0.08 M_{\odot} $ for Vela X-1, 
 $1.49 \pm 0.08 M_{\odot} $ for Cen X-3, $1.04 \pm 0.09 M_{\odot} $ for SMC X-1 and $1.073 \pm 0.36 M_{\odot} $ for Her X-1. We will restrict our discussions to these six stars only
  though there are similar data for other stars also.\\

\begin{figure}[htbp]
    \centering
        \includegraphics[scale=.3]{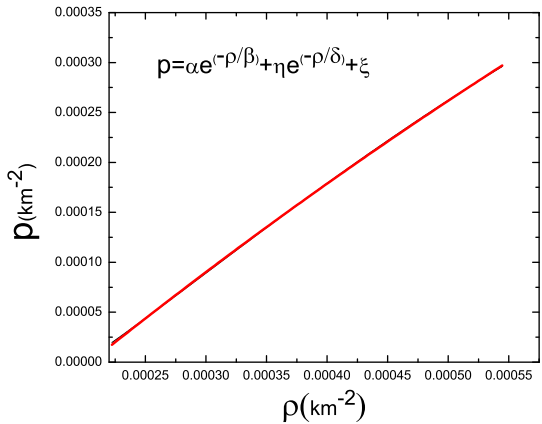}
        \caption{ Pressure ($p$) - Density($\rho$)(possible EOS) relation at the steller interior (taking a=0.002, C=0.52) 
        where $\alpha, \beta, \eta, \delta, \xi$ are constants and all are in units of $km^{-2}$. The pressure and density are also in units of $km^{-2}$.}
    \label{fig:11}
\end{figure}

In this article, we have investigated the physical behavior of the neutron
stars by considering it as isotropic pressure in nature and the space-time
of it to be described by metric Heint IIa\citep{H. Heintzmann1969}.\\
The observed masses and proposed radii for the neutron stars PSR J1614-2230, PSR J1903+327, Cen X-3, SMC X-1, Vela X-1, Her X-1 are used in this paper for an 
effective model of neutron stars.\\\\
We have obtained quite interesting results from this model, which are as follows:\\
\begin{enumerate}
 \item[(i)] The solutions are regular at the origin.
 \item[(ii)] Density and pressure variation at the interior of the neutron stars are well behaved(positive definite at the origin)[FIG.1 and FIG.2].
 \item[(iii)] Pressure reduced to zero at the boundary but density shows some finite value which is physically acceptable.
 \item[(iv)] In our model both the pressure and energy density are monotonically decreasing to the boundary.
 \item [(v)] It also satisfies TOV equation and energy conditions.
 \item[(vi)] Our isotropic neutron star model satisfy Herrera's stability condition \citep{Herrera1992}.
 \item[(vii)] From the mass-radius relation (9, 10), all desired interior features of a neutron stars can be evaluated.
 \item[(viii)] Our model satisfies Buchdahl mass-radius relation($\frac{ Mass}{Radius} < \frac{4}{9}$)\citep{H. A. Buchdahl1959}.
 \item[(ix)] According to our model the surface redshift of the neutron stars is found within standard value i.e. $Z_{s}\leq 0.85$ which is satisfactory\citep{Haensel2000}.
 \item[(x)] This model is stable with respect to infinitesimal radial perturbations.
 \item[(xi)] From FIG. 6 it is clear that our derived mass function is very much acceptable as the observed masses of various neutron stars are lying on the graph.
 \item[(xii)] We also give two possible equations of state(EOS) of matter at the interior of the star, one is polynomial(FIG. 9) 
 and another is exponential(FIG. 10). But we see that best fitted equation of state(EOS) is exponential.\\
 \item[(xiii)] We also estimated from our model that the equation of state of a neutron star would be like $ p = \alpha e^{(-\rho/\beta )} + \eta e^{(-\rho/\delta )} + \xi $ 
 where $\alpha, \beta,~ \eta,~ \delta, ~\xi$ are constant.
 \item[(xiv)] According to our model, equation of state (FIG 11) for a neutron star should not be soft equation of state rather it would be a stiff equation of state which 
 is also evidently proved in a result of F. $\ddot{O}$zel \cite{Ozel2006}.
\end{enumerate}

 Therefore our model is justified for the neutron star PSR J1614-2230, PSR J1903+327, Cen X-3, SMC X-1, Vela X-1 and Her X-1. Hence we can conclude with the hope that it may be helpful to 
 analyse the other neutron stars also. 

\section*{Acknowledgment} MK and SMH would like to thank IUCAA, Pune India and IMSc, Chennai, India for providing research facilities under Visiting Associateship.

\end{document}